\documentclass[journal]{IEEEtran}
\usepackage{amssymb}
\usepackage[cmex10]{amsmath}
\usepackage{stfloats}
\usepackage{graphicx}
\usepackage{subfigure}
\usepackage{epsfig,epsf,color,balance}
\usepackage{url}
\usepackage{bm}
\usepackage{mathrsfs}
\usepackage{amsthm}
\usepackage{algorithmic}

\usepackage[colorlinks]{hyperref}
\usepackage[applemac]{inputenc}

\newtheorem{theorem}{Theorem}

\usepackage{algorithm}
\usepackage{tensor}
\usepackage{graphicx}
\usepackage{epstopdf}
\IEEEoverridecommandlockouts
	
\begin{document}

\title{\LARGE Coding-Enhanced Cooperative Jamming for \\Secret Communication in Fluid Antenna Systems}

\author{
	\IEEEauthorblockN{Hao Xu, \emph{Member, IEEE}\IEEEauthorrefmark{0},    
		Kai-Kit Wong, \emph{Fellow, IEEE}\IEEEauthorrefmark{0},
		Wee Kiat New, \emph{Member, IEEE}\IEEEauthorrefmark{0},\\
		Guyue Li, \emph{Member, IEEE}\IEEEauthorrefmark{0},
		Farshad Rostami Ghadi, \emph{Member, IEEE}\IEEEauthorrefmark{0},
		Yongxu Zhu, \emph{Senior Member, IEEE}\IEEEauthorrefmark{0},\\
		Shi Jin, \emph{Fellow, IEEE}\IEEEauthorrefmark{0}
		Chan-Byoung Chae, \emph{Fellow, IEEE}\IEEEauthorrefmark{0} 
		and Yangyang Zhang
\vspace{-8mm}
	}
	
	\thanks{
	The work of H. Xu, K. K. Wong, W. K. New, and F. Rostami Ghadi is supported in part by the European Union's Horizon 2020 Research and Innovation Programme under MSCA Grant No. 101024636, and in part by the Engineering and Physical Sciences Research Council (EPSRC) under grant EP/W026813/1. The work of C. B. Chae is supported by the Institute of Information and Communication Technology Promotion (IITP) grant funded by the Ministry of Science and ICT (MSIT), Korea (No. 2021-0-02208, No. 2021-0-00486).}
	\thanks{H. Xu, K. K. Wong, W. K. New, and F. Rostami Ghadi are with the Department of Electronic and Electrical Engineering, University College London, London WC1E7JE, United Kingdom. K. K. Wong is also affiliated with Yonsei Frontier Laboratory, Yonsei University, Seoul, 03722, Korea (e-mail: $\rm \{hao.xu, kai\text{-}kit.wong, a.new, f.rostamighadi \}@ucl.ac.uk$). }
	\thanks{Guyue Li is with the School of Cyber Science and Engineering, Southeast University, Nanjing 210096, China. Guyue Li is also with Purple Mountain Laboratories, Nanjing 211111, China, and also with the Jiangsu Provincial Key Laboratory of Computer Network Technology, Nanjing 210096, China (e-mail: $\rm guyuelee@seu.edu.cn$).}
	\thanks{Y. Zhu and S. Jin are with the National Mobile Communications Research Laboratory, Southeast University, Nanjing 210096, China (e-mail: $\rm \{yongxu.zhu, jinshi\}@seu.edu.cn$).}
	\thanks{C. B. Chae is with the School of Integrated Technology, Yonsei University, Seoul, 03722, Korea (e-mail: $\rm cbchae@yonsei.ac.kr$).}
	\thanks{Y. Zhang is with Kuang-Chi Science Limited, Hong Kong SAR, China (e-mail: $\rm yangyang.zhang@kuang\text{-}chi.org$).}
	\thanks{Corresponding author: Hao Xu.}
}

\maketitle

\begin{abstract}
This letter investigates the secret communication problem for a fluid antenna system (FAS)-assisted wiretap channel, where the legitimate transmitter transmits an information-bearing signal to the legitimate receiver, and at the same time, transmits a jamming signal to interfere with the eavesdropper (Eve). Unlike the conventional jamming scheme, which usually transmits Gaussian noise that interferes not only with Eve but also with the legitimate receiver, in this letter, we consider that encoded codewords are transmitted to jam Eve. Then, by employing appropriate coding schemes, the legitimate receiver can successfully decode the jamming signal and then cancel the interference, while Eve cannot, even if it knows the codebooks. We aim to maximize the secrecy rate through port selection and power control. Although the problem is non-convex, we show that the optimal solution can be found. Simulation results show that by using the FAS technique and the proposed jamming scheme, the secrecy rate of the system can be significantly increased.
\end{abstract}


\IEEEpeerreviewmaketitle

\vspace{-2mm}
\section{Introduction}\label{section1}
\IEEEPARstart{R}{ecently}, the fluid antenna system (FAS) technology has emerged and shown promise for the next generation of wireless communication, e.g., \cite{wong2023fluid, ghadi2024physical}. FAS relies on flexible antenna technologies such as liquid-based antennas~\cite{huang2021liquid}, reconfigurable radio frequency (RF) pixel-based antennas~\cite{rodrigo2014frequency}, stepper motor-based antennas~\cite{basbug2017design}, etc. By adjusting the antenna position within a specified area, additional degrees of freedom and significant gains can be achieved~\cite{new2023information}. Unlike approaches such as massive multiple-input multiple-output (MIMO) or antenna selection techniques, which entail many RF chains fixed to antennas~\cite{chowdhury20206g, zhu2016geometric}, FAS offers a cost-effective way to take full advantage of the spatial channel variation.

Recently, the performance of FAS has been studied in different systems \cite{new2023fluid, ghadi2023copula, ma2023mimo, xu2023capacity, hu2024fluid, hu2024secure, tang2023fluid}. Specifically, \cite{new2023fluid} investigated the outage probability and diversity gain of a point-to-point FAS, where the transmitter and receiver used a conventional fixed-position antenna (FPA) and a fluid antenna, respectively. Exploiting the copula theory, \cite{ghadi2023copula} derived the closed-form expression of the outage probability under arbitrary correlated fading for a similar system. In \cite{ma2023mimo,xu2023capacity}, the capacity of a point-to-point and multi-access FAS-assisted system was respectively maximized, and it was shown that, in contrast to FPA, FAS could help greatly improve the system capacity. The total transmit power of a multi-access uplink system was minimized in~\cite{hu2024fluid}, where the base station has multiple fluid antennas. In~\cite{hu2024secure}, a wiretap channel consisting of a FAS-assisted legitimate transmitter (Alice), a legitimate receiver (Bob), and multiple eavesdroppers (Eves), was considered, and the worst secrecy rate was maximized by optimizing the antenna positions and beamforming vector at Alice. The secrecy problem for a FAS-assisted system has also been studied by \cite{tang2023fluid}. 
Different from \cite{hu2024secure}, in \cite{tang2023fluid}, only one Eve was considered, and Bob used FAS, while Alice and Eve used the traditional FPAs. In addition, besides the information-bearing signal, Alice also transmitted a jamming signal to interfere with Eve. The secrecy rate of the system was maximized by power control, and it was shown that by using FAS, the secrecy performance is comparable to the case where Bob uses multiple FPAs and maximal ratio combining (MRC). However, in \cite{tang2023fluid}, Alice jammed by simply transmitting Gaussian noise, which causes interference not only to Eve but also to Bob.

In this letter, we revisit the wiretap channel considered in \cite{tang2023fluid}. Different from \cite{tang2023fluid}, this letter assumes that Alice jams by transmitting encoded codewords such that by adopting appropriate coding approaches, Bob can successfully decode the jamming signal and then cancel the interference, while Eve cannot, even if it knows the codebooks. Using this jamming scheme, we aim to maximize the secrecy rate derived in \cite{xu2022new} by port selection and power control. Since the choices of the antenna position are discrete, the considered problem is a mixed-integer programming, which is NP-hard. However, we show that for any fixed antenna position, although the reduced problem is non-convex, the globally optimal solution can be obtained essentially in closed form, and this requires extremely low computational complexity. Therefore, the optimal solution of the original problem can be found by dealing with the reduced problem for all possible port selections. Simulation results show that in contrast to the conventional fixed antenna scheme, the secrecy rate of the system can be significantly increased by using FAS. In addition, the proposed jamming scheme can greatly outperform the method in \cite{tang2023fluid}.


\section{System Model and Problem Formulation}\label{GV_MAC-WT}
In this letter, we revisit the wiretap channel in \cite{tang2023fluid}, which consists of a legitimate transmitter (Alice), a legitimate receiver (Bob), and an Eve. Alice is equipped with two FPAs, one of which transmits an information-bearing signal to Bob and the other one transmits a jamming signal to interfere with Eve. Without loss of generality (w.l.o.g.), we assume that the first and second antennas of Alice, respectively, transmit the information-bearing signal $s_1 \sim {\cal CN}(0, p_1)$ and the jamming signal $s_2 \sim {\cal CN}(0, p_2)$, where $p_1$ and $p_2$ are the transmit power of the antennas. Eve has a single FPA. To enhance the secrecy performance, as in \cite{tang2023fluid}, we assume that Bob uses a linear FAS to receive the signal. Assume that the fluid antenna's location can be instantaneously switched to one of the $N$ predetermined ports, which are evenly distributed along a linear dimension of length $W \lambda$ and share a common RF chain.\footnote{This structure can be viewed as an approximation of an RF pixel-based linear FAS that has many compact antenna pixels, wherein a single port can be activated at a time by activating certain connections between the pixels \cite{new2023information}. This technology enables seamless port switching with virtually no time delay.} Here $W$ is the normalized size of the FAS and $\lambda$ is the wavelength. Then, the received signals at Bob and Eve are given by
\begin{align}\label{received_sig}
y^{(n)} & = h_1^{(n)} s_1 + h_2^{(n)} s_2 + \eta^{(n)}, \forall n \in \{1, \dots, N\}, \nonumber\\
z & = g_1 s_1 + g_2 s_2 + \mu, 
\end{align}
where $h_k^{(n)} \sim {\cal CN}(0, \sigma_k^2), \forall k \in \{1, 2\}$ is the channel gain from the $k$-th antenna of Alice to the $n$-th port of Bob, and $\eta^{(n)} \sim {\cal CN}(0, 1)$ is the additive white Gaussian noise (AWGN) experienced at the $n$-th port. Also, $g_1$, $g_2$, and $\mu$ are similarly defined for Eve. Note that the channel gains at different ports of a FAS are highly correlated. Denote $\bm h_k = [ h_k^{(1)}, \dots, h_k^{(N)} ]^T$ and its covariance matrix by $\bm \varSigma_k = \sigma_k^2 \bm \varSigma$. We characterize the spatial correlation of the ports by following the Jake's model so that the $(n,n')$-th element of $\bm \varSigma$ is given by \cite{hao2023on, xu2024revisiting}
\begin{equation}\label{mn}
(\bm \varSigma)_{n,n'} = J_0 \left( 2 \pi (n-n') \varDelta \right),
\end{equation}
where $\varDelta = W / (N-1)$ is the normalized distance between any two adjacent ports and $J_0 (\cdot)$ is the zero-order Bessel function of the first kind.

A popular jamming scheme is to use a cooperative jammer, which is the second antenna of Alice in our case, to simply transmit Gaussian noise. In this case, the secrecy rate is
\begin{align}\label{rate_GN}
	& R_{\text {GN}}^{(n)} (p_1, p_2) \nonumber\\
	= & \left[ \log\! \left( 1 \!+\! \frac{ p_1 | h_1^{(n)} |^2 }{ p_2 | h_2^{(n)} |^2 \!+\! 1 } \right) - \log\! \left( 1 \!+\! \frac{ p_1 | g_1 |^2 }{ p_2 | g_2 |^2 \!+\! 1 } \right) \right]^+\!\!,
\end{align}
where $[a]^+=\max\{0,a\}$ and the subscript ``GN'' stands for Gaussian noise. The problem of maximizing (\ref{rate_GN}) by port selection and power control has been studied in \cite{tang2023fluid}. It can be easily found from (\ref{rate_GN}) that if Gaussian noise is transmitted, the jamming signal interferes not only with Bob but also with Eve. To avoid the jamming signal from interfering with Bob, it has been shown in \cite{xu2022new} that the coding-enhanced jamming scheme can be used, i.e., the second antenna of Alice jams by transmitting encoded codewords instead of Gaussian noise. By adopting appropriate coding approaches, Bob can successfully decode the jamming signal and then cancel the interference, while Eve cannot, even if it knows the codebooks. In this case, an achievable secrecy rate is given by \cite[Theorem~1]{xu2022new}
\begin{equation}\label{rate_EJ}
R_{\text {EJ}}^{(n)} (p_1, p_2) = \max \{ \min \{ {\hat R}, {\tilde R} \}, {\bar R} \},
\end{equation}
where the subscript ``EJ'' indicates that the second antenna of Alice is acting as an encoded jammer, and ${\hat R}$, ${\tilde R}$, and ${\bar R}$ are, respectively, given by
\begin{align}\label{R_hat_tilde_bar}
	{\hat R} & = \left[ \log \left( 1 + p_1 | h_1^{(n)} |^2 \right) - \log \left( 1 + \frac{ p_1 | g_1 |^2 }{ p_2 | g_2 |^2 + 1 } \right) \right]^+, \nonumber\\
	{\tilde R} & = \!\left[\! \log\! \left(\! 1 \!+\! p_1 | h_1^{\!(n)} \!|^2 \!+\! p_2 | h_2^{\!(n)} \!|^2 \!\right) \!-\! \log\! \left( 1 \!+\! p_1 | g_1 |^2 \!+\! p_2 | g_2 |^2 \right) \!\right]^{\!+}\!\!\!, \nonumber\\
	{\bar R} & = \left[ \log \left( 1 + p_1 | h_1^{(n)} |^2 \right) - \log \left( 1 + p_1 | g_1 |^2 \right) \right]^+.
\end{align}
From (\ref{rate_GN})$\sim$(\ref{R_hat_tilde_bar}), we recognize that $R_{\text {EJ}}^{(n)} (p_1, p_2)$ is quite different from $R_{\text {GN}}^{(n)} (p_1, p_2)$. According to \cite{xu2022new}, to apply the coding-enhanced jamming scheme, besides the secret message, Alice also has to transmit auxiliary messages at both antennas at some rate. ${\hat R}$ and ${\tilde R}$ are, respectively, the upper bounds to the rate of the secret message and the sum rate of all messages transmitted by Alice. Then, $\min \{ {\hat R}, {\tilde R} \}$ is an achievable secrecy rate. Since the rate ${\bar R}$ is always achievable, $R_{\text {EJ}}^{(n)} (p_1, p_2)$ takes on the larger of $\min \{ {\hat R}, {\tilde R} \}$ and ${\bar R}$.

We maximize $R_{\text {EJ}}^{(n)} (p_1, p_2)$ by port selection and power control, i.e., considering the following problem:
\begin{align}\label{max_SR_EJ}
	\mathop {\max }\limits_{ n, p_1, p_2 } \quad & R_{\text {EJ}}^{(n)} (p_1, p_2)  \nonumber\\
	\text{s.t.} \quad\; & n \in \{1, \dots, N\}, \nonumber\\
	& p_1, p_2 \geq 0, ~  p_1 + p_2 \leq P,
\end{align}
where $P$ is the total transmit power constraint of Alice.

\section{Main Results}
Problem (\ref{max_SR_EJ}) is a mixed-integer programming, which is NP-hard. To address this issue, we fix $n$ and consider the following simplified problem:
\begin{align}\label{max_SR_EJ1}
	\mathop {\max }\limits_{ p_1, p_2 } \quad & R_{\text {EJ}}^{(n)} (p_1, p_2)  \nonumber\\
	\text{s.t.} \quad\; & p_1, p_2 \geq 0, ~  p_1 + p_2 \leq P.
\end{align}
In the following, we show that the optimal solution of (\ref{max_SR_EJ1}) in closed form can be obtained. 

It is obvious from (\ref{rate_EJ}) that (\ref{max_SR_EJ1}) can be solved by separately maximizing $\min \{ {\hat R}, {\tilde R} \}$ and ${\bar R}$. From (\ref{R_hat_tilde_bar}), we know that ${\bar R}$ can be obtained directly from ${\hat R}$ and ${\tilde R}$ by simply setting $p_2 = 0$. Therefore, problem (\ref{max_SR_EJ1}) is equivalent to
\begin{align}\label{max_SR_EJ2}
	\mathop {\max }\limits_{ p_1, p_2 } \quad & \min \{ {\hat R}, {\tilde R} \}  \nonumber\\
	\text{s.t.} \quad\; & p_1, p_2 \geq 0, ~  p_1 + p_2 \leq P,
\end{align}
which is a max-min problem and is usually intractable. However, we show below that it can be solved optimally. To derive the result, we first consider the following two problems:
\begin{align}
	\mathop {\max }\limits_{ p_1, p_2 } \quad & {\hat R}  \nonumber\\
	\text{s.t.} \quad\; & {\hat P}_{1, {\text {lb}}} \leq p_1 \leq {\hat P}_{1, {\text {ub}}}, ~ p_2 \geq 0, ~ p_1 + p_2 \leq P, \label{max_R_hat}\\
	\mathop {\max }\limits_{ p_1, p_2 } \quad & {\tilde R}  \nonumber\\
	\text{s.t.} \quad\; & {\tilde P}_{1, {\text {lb}}} \leq p_1 \leq {\tilde P}_{1, {\text {ub}}}, ~ p_2 \geq 0, ~ p_1 + p_2 \leq P, \label{max_R_tilde}
\end{align}
where ${\hat P}_{1, {\text {lb}}}$, ${\hat P}_{1, {\text {ub}}}$, ${\tilde P}_{1, {\text {lb}}}$, and ${\tilde P}_{1, {\text {ub}}}$ are new lower and upper bounds set to $p_1$. Their values will be specified in Theorem~\ref{theorem2}. From the expressions of ${\hat R}$ and ${\tilde R}$ in (\ref{R_hat_tilde_bar}), it is known that both (\ref{max_R_hat}) and (\ref{max_R_tilde}) are non-convex. However, in the following Theorem~\ref{theorem1}, we show that both of them can be optimally solved. To make it easy to distinguish, we use $({\hat p}_1^*, {\hat p}_2^*)$ and $({\tilde p}_1^*, {\tilde p}_2^*)$ to, respectively, denote the optimal solutions of (\ref{max_R_hat}) and (\ref{max_R_tilde}).

\begin{theorem}\label{theorem1}
The optimal solution of (\ref{max_R_hat}) is given by
\begin{equation}\label{optimal_p1}
		{\hat p}_1^* = \left\{\!\!\!\!\!\!\!\!
		\begin{array}{ll}
			& - \frac{c}{b}, ~{\text {if}}~ a = 0 ~ {\text {and}}~ {\hat P}_{1, {\text {lb}}} < - \frac{c}{b} < {\hat P}_{1, {\text {ub}}}, \vspace{0.5 em}\\
			& \arg \mathop {\max }\limits_{p_1 \in \{ \alpha, {\hat P}_{1, {\text {ub}}}\}} {\hat R}, ~{\text {if}}~ a > 0 ~ {\text {and}}~ {\hat P}_{1, {\text {lb}}} < \alpha < {\hat P}_{1, {\text {ub}}}, \vspace{0.5 em}\\
			& \arg \mathop {\max }\limits_{p_1 \in \{ {\hat P}_{1, {\text {lb}}}, \alpha\}} {\hat R}, ~~{\text {if}}~ a < 0,~ b^2 - 4ac > 0,\\
			& \quad\quad\quad\quad\quad\quad\quad\quad {\text {and}}~ {\hat P}_{1, {\text {lb}}} < \alpha < {\hat P}_{1, {\text {ub}}}, \vspace{0.5 em}\\
			& \arg \mathop {\max }\limits_{p_1 \in \{ {\hat P}_{1, {\text {lb}}}, {\hat P}_{1, {\text {ub}}}\}} {\hat R}, ~~{\text {otherwise}}, \\
		\end{array} \right.
\end{equation}
and ${\hat p}_2^* = P - {\hat p}_1^*$, where $\alpha = \frac{-b - \sqrt{b^2 - 4ac}}{2 a}$ and
\begin{align}\label{abc}
		a & = - | h_1^{(n)} |^2 | g_2 |^2 (| g_1 |^2 - | g_2 |^2), \nonumber\\
		b & = - 2 | h_1^{(n)} |^2 | g_2 |^2 (P | g_2 |^2 + 1), \nonumber\\
		c & = | h_1^{(n)} |^2 (P | g_2 |^2 + 1)^2 - | g_1 |^2 (P | g_2 |^2 + 1).
\end{align}
On the other hand, the optimal solution $({\tilde p}_1^*, {\tilde p}_2^*)$ of (\ref{max_R_tilde}) takes the value of $({\tilde P}_{1, {\text {lb}}}, 0)$, $({\tilde P}_{1, {\text {lb}}}, P - {\tilde P}_{1, {\text {lb}}})$, $({\tilde P}_{1, {\text {ub}}}, 0)$, or $({\tilde P}_{1, {\text {ub}}}, P - {\tilde P}_{1, {\text {ub}}})$, which maximizes ${\tilde R}$, i.e.,	
	\begin{align}\label{opt_power_R_tilde}
		& ({\tilde p}_1^*, {\tilde p}_2^*) = \nonumber\\
		& \arg \mathop {\max }\limits_{ (p_1, p_2) \in \left\{ ({\tilde P}_{1, {\text {lb}}}, 0), ({\tilde P}_{1, {\text {lb}}}, P - {\tilde P}_{1, {\text {lb}}}), ({\tilde P}_{1, {\text {ub}}}, 0), ({\tilde P}_{1, {\text {ub}}}, P - {\tilde P}_{1, {\text {ub}}}) \right\} } {\tilde R}.
\end{align}
\end{theorem}

\itshape \textbf{Proof:}  \upshape
See Appendix \ref{prove_theorem1}.
\hfill $\Box$

\begin{algorithm}[t]
	\caption{Algorithm for solving (\ref{max_SR_EJ})}
	\label{algorithm1}
	\begin{algorithmic}[1]
		\FOR{$n = 1:N$}
		\IF{$| g_2 |^2 \leq | h_2^{(n)} |^2 ( 1 + P | h_1^{(n)} |^2 )^{-1}$}
		\STATE Let $({\hat P}_{1, {\text {lb}}}, {\hat P}_{1, {\text {ub}}}) = (0, P)$, solve (\ref{max_R_hat}) by Theorem~\ref{theorem1}, obtain $({\hat p}_1^*, {\hat p}_2^*)$, and compute $R_{\text {EJ}}^{(n)*} = R_{\text {EJ}}^{(n)} ({\hat p}_1^*, {\hat p}_2^*)$.
		\ELSIF{$| g_2 |^2 \geq | h_2^{(n)} |^2$}
		\STATE Let $({\tilde P}_{1, {\text {lb}}}, {\tilde P}_{1, {\text {ub}}}) = (0, P)$, solve (\ref{max_R_tilde}) by Theorem~\ref{theorem1}, obtain $({\tilde p}_1^*, {\tilde p}_2^*)$, and compute $R_{\text {EJ}}^{(n)*} = R_{\text {EJ}}^{(n)} ({\tilde p}_1^*, {\tilde p}_2^*)$.
		\ELSE
		\STATE Let $({\hat P}_{1, {\text {lb}}}, {\hat P}_{1, {\text {ub}}}) = (0, \beta)$ and $({\tilde P}_{1, {\text {lb}}}, {\tilde P}_{1, {\text {ub}}}) = (\beta, P)$, and solve (\ref{max_R_hat}) and (\ref{max_R_tilde}) by Theorem~\ref{theorem1}. Compute $R_{\text {EJ}}^{(n)*} = \max \{R_{\text {EJ}}^{(n)} ({\hat p}_1^*, {\hat p}_2^*), R_{\text {EJ}}^{(n)} ({\tilde p}_1^*, {\tilde p}_2^*)\}$.
		\ENDIF
		\ENDFOR
		\STATE The optimal solution of (\ref{max_SR_EJ}) is $\max \{R_{\text {EJ}}^{(1)*}, \dots, R_{\text {EJ}}^{(N)*}\}$.
	\end{algorithmic}
\end{algorithm}

Based on Theorem~\ref{theorem1}, (\ref{max_SR_EJ2}) can be optimally solved and the results are provided in the following theorem.

\begin{theorem}\label{theorem2}
Problem (\ref{max_SR_EJ2}) can be optimally solved by discussing three different cases as follows.
\begin{itemize}
		\item If $| g_2 |^2 \leq | h_2^{(n)} |^2 ( 1 + P | h_1^{(n)} |^2 )^{-1}$,	problem (\ref{max_SR_EJ2}) reduces to (\ref{max_R_hat}) with $({\hat P}_{1, {\text {lb}}}, {\hat P}_{1, {\text {ub}}}) = (0, P)$. 
		Its optimal solution is thus $({\hat p}_1^*, {\hat p}_2^*)$, which can be obtained based on (\ref{optimal_p1}).
		\item If $| g_2 |^2 \geq | h_2^{(n)} |^2$,	problem (\ref{max_SR_EJ2}) reduces to (\ref{max_R_tilde}) with $({\tilde P}_{1, {\text {lb}}}, {\tilde P}_{1, {\text {ub}}}) = (0, P)$. 
		Its optimal solution is thus $({\tilde p}_1^*, {\tilde p}_2^*)$, which can be obtained based on (\ref{opt_power_R_tilde}).
		\item If $| h_2^{(n)} |^2 ( 1 + P | h_1^{(n)} |^2 )^{-1} < | g_2 |^2 < | h_2^{(n)} |^2$, problem (\ref{max_SR_EJ2}) can be divided into two subproblems, i.e., (\ref{max_R_hat}) with $({\hat P}_{1, {\text {lb}}}, {\hat P}_{1, {\text {ub}}}) = (0, \beta)$ and (\ref{max_R_tilde}) with $({\tilde P}_{1, {\text {lb}}}, {\tilde P}_{1, {\text {ub}}}) = (\beta, P)$, where $\beta = (| h_2^{(n)} |^2 / | g_2 |^2 - 1) / | h_1^{(n)} |^2$.
		Then, the optimal solution of (\ref{max_SR_EJ2}) is $({\hat p}_1^*, {\hat p}_2^*)$ if ${\hat R} ({\hat p}_1^*, {\hat p}_2^*) \geq {\tilde R} ({\tilde p}_1^*, {\tilde p}_2^*)$, and $({\tilde p}_1^*, {\tilde p}_2^*)$ otherwise.
\end{itemize}
\end{theorem}

\itshape \textbf{Proof:}  \upshape
See Appendix \ref{prove_theorem2}.
\hfill $\Box$

Based on Theorem~\ref{theorem2}, the optimal solution of (\ref{max_SR_EJ1}) or (\ref{max_SR_EJ2}) can be obtained by discussing different values of $| g_2 |^2$. Since there are only three cases, and in each case, the optimal solution of the reduced problem (\ref{max_R_hat}) or (\ref{max_R_tilde}) can be obtained from Theorem~\ref{theorem1} in closed-form, it requires quite low computational complexity to solve (\ref{max_SR_EJ1}). Therefore, the optimal solution of (\ref{max_SR_EJ}) can be found by dealing with (\ref{max_SR_EJ1}) for all possible $n \in \{1, \dots, N\}$. The details are summarized in Algorithm~\ref{algorithm1}.


\section{Simulation Results}\label{simul}

\begin{figure}
	\centering
	\includegraphics[scale=0.44]{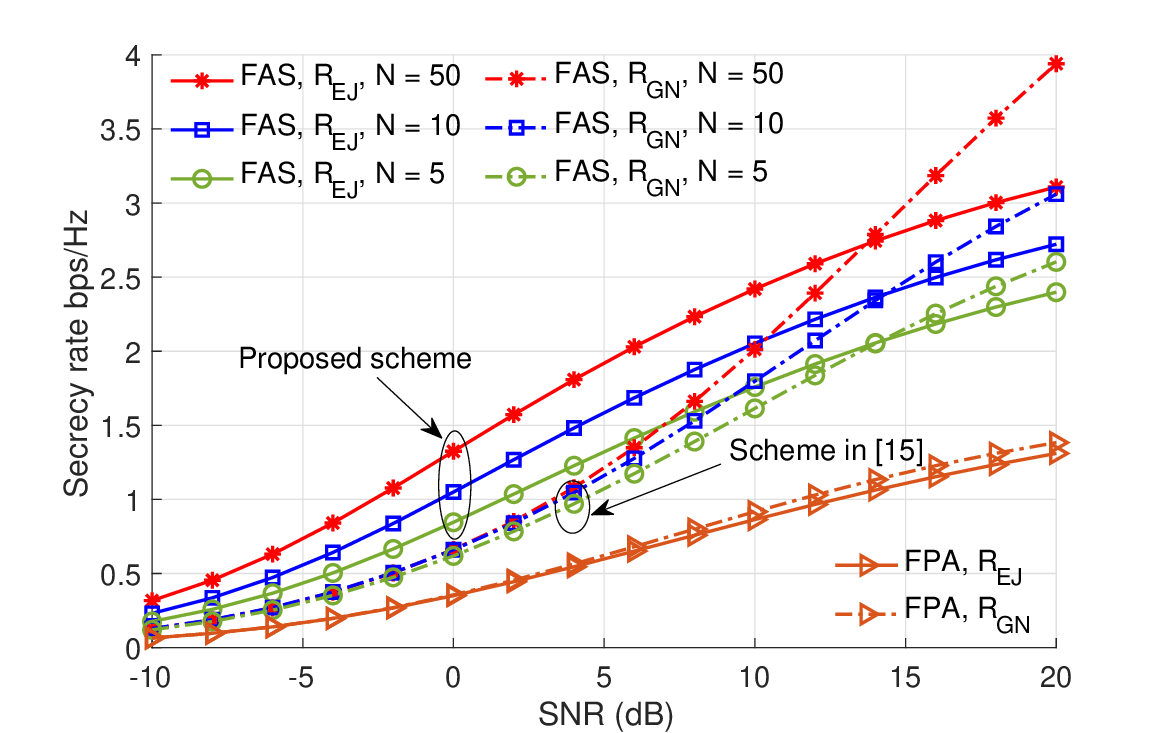}
	\vspace{-1 em}
	\caption{Secrecy rate $R_{\text {EJ}}$ and $R_{\text {GN}}$ versus the SNR $\rho$ with $W = 5$.}\label{SR_VS_P}
	\vspace{-1.5 em}
\end{figure}

In this section, we evaluate the performance of the proposed jamming scheme by simulation. Define the signal-to-noise ratio (SNR) as $\rho = 10 \log_{10} P$ dB. All simulation results are obtained by averaging over $10^5$ channel realizations. In each realization, the channel gains $g_1$ and $g_2$ are generated according to independent and identically distributed (i.i.d.) complex Gaussian distribution with zero mean and unit variance. Differently, the elements of $\bm h_k$ are correlated. To ensure that $h_k^{(n)} \sim {\cal CN}(0, \sigma_k^2)$ and ${\mathbb E} \left[ \bm h_k \bm h_k^H \right] = \sigma_k^2 \bm \varSigma$, we generate the channel vector $\bm h_k$ by using (\ref{mn}) and the technique in \cite{xu2024revisiting}. In particular, let $\bm U \bm \varTheta \bm U^H$ denote the eigen-decomposition of $\bm \varSigma$ and $\bm h_k = \sigma_k \bm U \bm \varTheta^{\frac{1}{2}} \bm x_k$, where $\bm x_k = [ x_k^{(1)}, \dots, x_k^{(N)} ]^T$ and $x_k^{(n)}, \forall n \in \{1, \dots, N\}$ follow i.i.d. ${\cal CN}(0, 1)$. It can be seen that $\bm h_k$ generated above satisfies the distribution model.

For comparison, the results obtained by the Gaussian noise jamming method in \cite{tang2023fluid}, the conventional FPA scheme, which can be seen as a special FAS case with $N = 1$, and the equal power allocation method are depicted as benchmarks. Note that although we obtain the achievable secrecy rate (\ref{rate_EJ}) based on \cite[Theorem 1]{xu2022new}, a direct comparison of the results between this letter and \cite{xu2022new} is not available for two reasons. First, in this letter, Bob applies FAS, while in \cite{xu2022new}, Bob uses the conventional FPA. Second, this letter assumes that Alice has two transmitting antennas, one of which transmits the information-bearing signal and the other transmits the jamming signal. These two signals have the maximum sum power constraint. Differently, in \cite{xu2022new}, the information-bearing and jamming signals are from two different transmitters, and thus have independent maximum power constraint. These differences make it impossible to solve the considered problem by directly using the scheme in \cite{xu2022new}.

\begin{figure}
	\centering
	\includegraphics[scale=0.44]{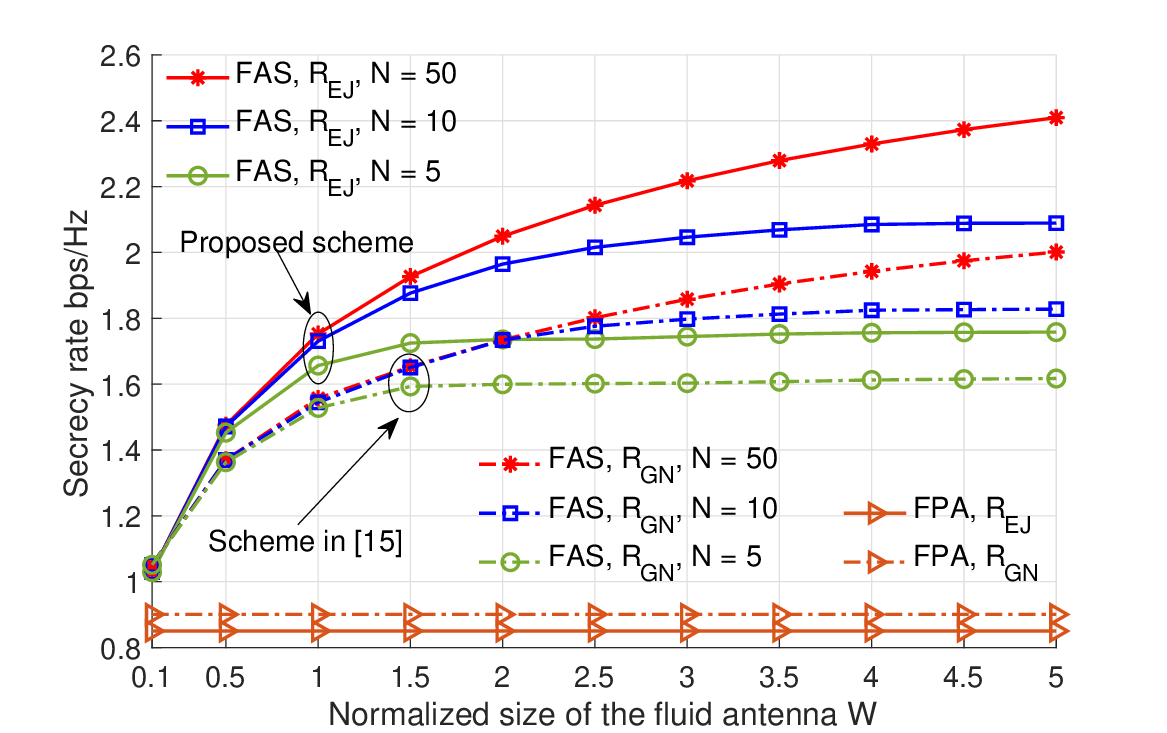}
	\vspace{-0.5 em}
	\caption{Secrecy rate $R_{\text {EJ}}$ and $R_{\text {GN}}$ versus $W$ with $\rho = 10$ dB.}\label{SR_VS_W}
	\vspace{-1 em}
\end{figure}

\begin{figure}
	\centering
	\includegraphics[scale=0.44]{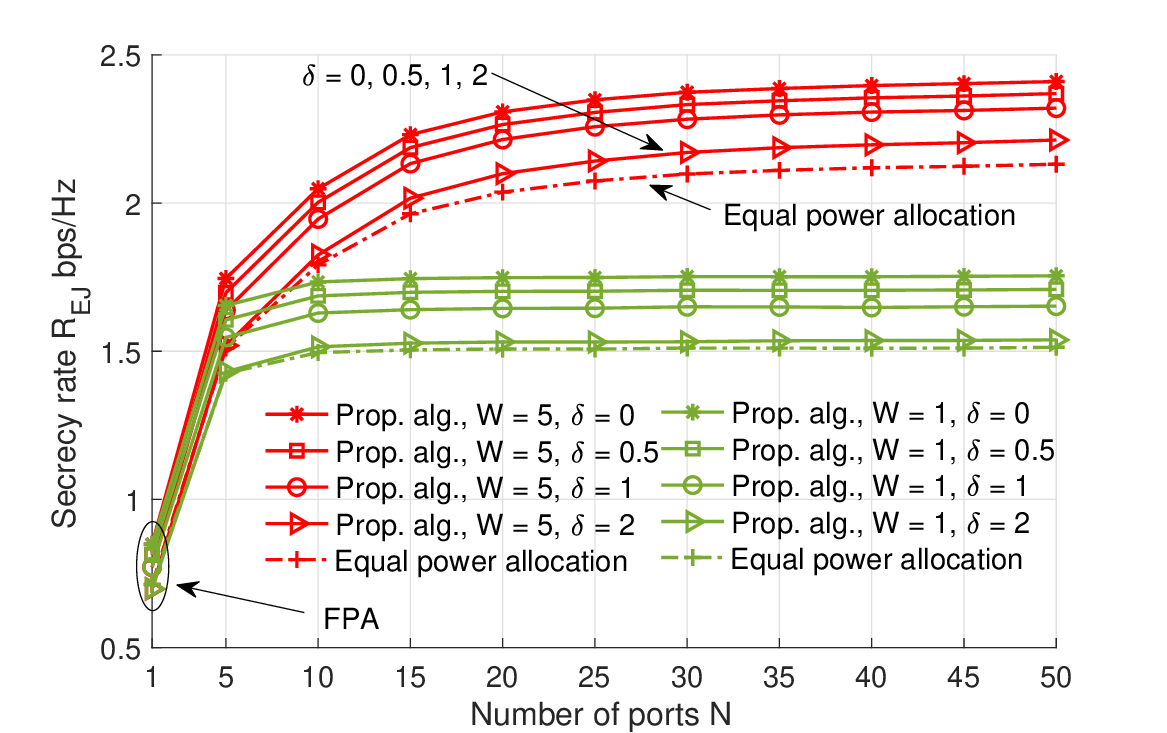}
	\vspace{-0.5 em}
	\caption{Secrecy rate $R_{\text {EJ}}$ versus $N$ with $\rho = 10$ dB.}\label{SR_VS_N}
	\vspace{-1.5 em}
\end{figure}

Fig.~\ref{SR_VS_P} depicts the secrecy rate obtained by different schemes versus $\rho$. Several observations can be made from this figure. First, as expected, the secrecy rate obtained by all schemes increases with $\rho$. Second, no matter which jamming scheme is applied, compared with FPA, using FAS can greatly improve the secrecy performance of the system. Moreover, Fig.~\ref{SR_VS_P} shows that when FAS is used and $\rho$ is relatively small, compared with the method in \cite{tang2023fluid}, the secrecy rate of the system can be greatly increased by the proposed coding-enhanced jamming scheme. For example, when $N = 50$ and $\rho = 5$, an increase in the secrecy rate of over $50\%$ can be obtained. However, when $\rho$ is large, the method in \cite{tang2023fluid} has a better secrecy performance.

In Fig.~\ref{SR_VS_W}, the effect of the normalized size of FAS $W$ is investigated. It can be seen that the secrecy performance of the system can be greatly improved by using FAS. We also see that for a given $N$, when $W$ increases, the secrecy rate obtained by both the proposed scheme and the method in \cite{tang2023fluid} significantly increases at the beginning since increasing $W$ reduces the correlation among ports. However, when $W$ becomes large, the secrecy rate saturates. This is because if $W$ is large enough, there will be almost no spatial correlation among ports. For example, when $N = 10$ and $W = 4.5$, the distance between any two adjacent ports is $\lambda/2$, under which the channel gains observed at different ports are usually considered to be independent. Then, further increasing $W$ for a fixed $N$ does not bring any gain in the secrecy rate.

For the sake of simplicity, this letter assumes perfect CSI of all links. However, it is actually not easy to obtain the perfect CSI between Alice and Eve. In Fig.~\ref{SR_VS_N}, we investigate the secrecy performance of the system in the case where only partial CSI of the Alice-Eve link is known. Specifically, as in \cite{tang2023fluid}, we assume $g_k = {\hat g}_k + \varDelta_k, \forall k \in \{1, 2\}$, where ${\hat g}_k$ and $\varDelta_k$ are, respectively, the estimated channel and estimation error. Also, $\varDelta_k$ is complex and uniformly distributed in a circular region of radius $\delta |g_k|$, where $\delta$ is the normalized channel uncertainty. Alice knows perfect $\bm h_k$, but only the estimate ${\hat g}_k$, based on which the proposed algorithm is performed. For comparison, we also depict the secrecy rate obtained by equal power allocation, i.e., $p_1 = p_2 = P/2$. Using this scheme, no CSI of the Alice-Eve link is needed. As expected, Fig.~\ref{SR_VS_N} shows that the secrecy rate reduces with $\delta$. Moreover, we see that although Alice only knows imperfect CSI, the performance obtained by using the proposed algorithm still outperforms the equal power allocation scheme. Fig.~\ref{SR_VS_N} also shows that for a given $W$, the secrecy rate first increases greatly with $N$ and then saturates. This is because the antenna ports are highly correlated. Increasing $N$ initially helps enhance the secrecy by introducing additional diversity. But as $N$ becomes large, the benefits of increasing $N$ diminish due to the stronger inter-correlation resulting from smaller port distances.


\section{Conclusions}\label{section6}
This letter revisited the FAS-assisted wiretap channel considered in \cite{tang2023fluid}, where one of the antennas of Alice plays as a cooperative jammer. To avoid the jamming signal from interfering with Bob, we assumed that Alice jams by transmitting encoded codewords instead of Gaussian noise as in \cite{tang2023fluid}. We maximized the secrecy rate by port selection and power control. Despite non-convexity, we proved that the optimal solution could be found. Simulation results confirmed the superior performance of FAS over the traditional FPA system, and also that of the proposed jamming scheme over the method provided in \cite{tang2023fluid}.
Note that to be consistent with the model in \cite{tang2023fluid}, we assumed that Eve is equipped with the conventional FPA. In the future, we will consider the more general case where Eve also uses FAS to maximize its benefit.

\appendices

\section{Proof of Theorem~\ref{theorem1}}\label{prove_theorem1}
It is obvious from (\ref{R_hat_tilde_bar}) that in the optimal case of (\ref{max_R_hat}), the constraint $p_1 + p_2 \leq P$ should always hold with equality since otherwise, the objective function ${\hat R}$ can be further increased by increasing $p_2$. Letting $p_2 = P - p_1$, ${\hat R}$ in (\ref{R_hat_tilde_bar}) can be reformulated as 
\begin{align}
	{\hat R} (p_1) & = \log \left( 1 + p_1 | h_1^{(n)} |^2 \right) - \log \left( 1 + \frac{ p_1 | g_1 |^2 }{ (P - p_1) | g_2 |^2 + 1 } \right) \nonumber\\
	& = \log \left( 1 + p_1 | h_1^{(n)} |^2 \right) + \log \left( P | g_2 |^2 + 1 - p_1 | g_2 |^2 \right) \nonumber\\
	& - \log \left[ P | g_2 |^2 + 1 + p_1 (| g_1 |^2 - | g_2 |^2) \right].
\end{align}
Its first-order derivative over $p_1$ is
\begin{align}\label{deri}
	& \partial {\hat R} / \partial p_1 = \nonumber\\
	& \frac{(a p_1^2 + b p_1 + c) / \ln 2}{( 1 \!\!+\!\! p_1 | h_1^{(n)} |^2 )  \!\left( P | g_2 |^2  \!\!+\!\! 1 \!\!- \!\! p_1 | g_2 |^2 \right)  \!\left[ P | g_2 |^2  \!\!+ \!\! 1  \!\!+ \!\! p_1 (| g_1 |^2  \!\!- \!\! | g_2 |^2) \right] },
\end{align}
where $a$, $b$, and $c$ are defined in (\ref{abc}). It is seen that regardless of the value of $p_1$, the denominator of (\ref{deri}) is always positive. In the following, we analyze the monotonicity of ${\hat R}$ over $p_1$ and obtain (\ref{optimal_p1}) by discussing the values of $a$, $b$, and $c$.

{\bf Case~$1$:}
First, we consider the scenario with $a = 0$. In this case, $\partial {\hat R} / \partial p_1$ has zero point $- \frac{c}{b}$. It is obvious from (\ref{abc}) that $b < 0$. Then, if ${\hat P}_{1, {\text {lb}}} < - \frac{c}{b} < {\hat P}_{1, {\text {ub}}}$, ${\hat R}$ increases with $p_1$ in $[ {\hat P}_{1, {\text {lb}}}, - \frac{c}{b} ]$ and decreases in $[ - \frac{c}{b}, {\hat P}_{1, {\text {ub}}} ]$. In the other cases, e.g., $- \frac{c}{b} \leq {\hat P}_{1, {\text {lb}}}$ or $- \frac{c}{b} \geq {\hat P}_{1, {\text {ub}}}$, it can be seen that ${\hat R}$ either decreases or increases with $p_1$ in the whole feasible region $[ {\hat P}_{1, {\text {lb}}}, {\hat P}_{1, {\text {ub}}} ]$. Hence, the optimal $p_1$ is either ${\hat P}_{1, {\text {lb}}}$ or ${\hat P}_{1, {\text {ub}}}$.

{\bf Case~$2$:}
Second, we discuss the case with $a > 0$. According to (\ref{abc}) and using the fact that $b < 0$, we have
\begin{equation}\label{in_root}
	b^2 - 4ac = b^2 \frac{| g_1 |^2}{| g_2 |^2} - 2ab \frac{| g_1 |^2}{| h_1^{(n)} |^2 | g_2 |^2} > 0.
\end{equation}
Then, the numerator of (\ref{deri}), i.e., $a p_1^2 + b p_1 + c$, is an upward parabola and has the following two zero points
\begin{align}\label{zero_point}
	\alpha = \frac{-b - \sqrt{b^2 - 4ac}}{2 a},~ \alpha' = \frac{-b + \sqrt{b^2 - 4ac}}{2 a},
\end{align}
with $\alpha < \alpha'$. By comparing the values of the two zero points with those of the boundary points ${\hat P}_{1, {\text {lb}}}$ and ${\hat P}_{1, {\text {ub}}}$, the monotonicity of ${\hat R}$ over $p_1$ can be completely revealed and the optimal $p_1$ can thus be obtained. However, to this end, we need to discuss many different cases. To be concise, we demonstrate that the optimal $p_1$ can be readily determined by checking the value of ${\hat R} (p_1)$ at only a few possible solutions.

If ${\hat P}_{1, {\text {lb}}} \leq \alpha \leq {\hat P}_{1, {\text {ub}}}$, $\partial {\hat R} / \partial p_1$ is non-negative in $[ {\hat P}_{1, {\text {lb}}}, \alpha ]$. Therefore, ${\hat R}$ increases with $p_1$ in $[ {\hat P}_{1, {\text {lb}}}, \alpha ]$. The monotonicity of ${\hat R}$ in the remaining feasible region $[ \alpha, {\hat P}_{1, {\text {ub}}} ]$ depends on whether ${\hat P}_{1, {\text {ub}}}$ is greater than the second zero point $\alpha'$ or vice versa. However, regardless of the circumstances, it can be easily checked that the optimal $p_1$ is either $\alpha$ or ${\hat P}_{1, {\text {ub}}}$. If the condition ${\hat P}_{1, {\text {lb}}} \leq \alpha \leq {\hat P}_{1, {\text {ub}}}$ does not hold, although there are many possible monotonicity cases, it can be readily verified that the optimal $p_1$ is either ${\hat P}_{1, {\text {lb}}}$ or ${\hat P}_{1, {\text {ub}}}$. Due to space limitation, we omit the details here.

{\bf Case~$3$:}
Next, we consider the case with $a<0$.
If in this case we further have $b^2 - 4ac > 0$, it is obvious that $\alpha$ and $\alpha'$ are also zero points of $a p_1^2 + b p_1 + c$, but different from Case~$2$, we have $\alpha > \alpha'$ since $a$ is negative. It can be similarly proven as Case~$2$ that if ${\hat P}_{1, {\text {lb}}} \leq \alpha \leq {\hat P}_{1, {\text {ub}}}$, the optimal $p_1$ is either ${\hat P}_{1, {\text {lb}}}$ or $\alpha$. In addition, if one or both of the conditions $b^2 - 4ac > 0$ and ${\hat P}_{1, {\text {lb}}} \leq \alpha \leq {\hat P}_{1, {\text {ub}}}$ are not true, the optimal $p_1$ is either ${\hat P}_{1, {\text {lb}}}$ or ${\hat P}_{1, {\text {ub}}}$.

Now we prove (\ref{opt_power_R_tilde}). ${\tilde R}$ in (\ref{R_hat_tilde_bar}) can be rewritten as 
\begin{align}\label{R_tilde}
	{\tilde R} & = \log \left( 1 + p_1 \frac{| h_1^{(n)} |^2}{1 + p_2 | h_2^{(n)} |^2} \right) - \log \left( 1 + p_1 \frac{| g_1 |^2}{1 + p_2 | g_2 |^2} \right) \nonumber\\
	& + \log \left( 1 + p_2 | h_2^{(n)} |^2 \right) - \log \left( 1 + p_2 | g_2 |^2 \right),
\end{align}
from which it can be seen that for any given $p_2$, the optimal $p_1$ that maximizes ${\tilde R}$ is either ${\hat P}_{1, {\text {lb}}}$ or ${\hat P}_{1, {\text {ub}}}$, i.e., ${\hat p}_1^* = {\hat P}_{1, {\text {lb}}}$ if 
\begin{equation}
	\frac{| h_1^{(n)} |^2}{1 + p_2 | h_2^{(n)} |^2} \leq \frac{| g_1 |^2}{1 + p_2 | g_2 |^2},
\end{equation}
and ${\hat p}_1^* = {\hat P}_{1, {\text {ub}}}$ vice versa. Similarly, it can be easily proven that for any $p_1$,  the optimal $p_2$ that maximizes ${\tilde R}$ is either $0$ or $P - p_1$. As a result, (\ref{opt_power_R_tilde}) is thus proven. This completes the proof of Theorem~\ref{theorem1}.

\section{Proof of Theorem~\ref{theorem2}}\label{prove_theorem2}
Using (\ref{R_hat_tilde_bar}) and neglecting the $[ \cdot ]^+$ operation, the difference between ${\hat R}$ and ${\tilde R}$ is given by
\begin{align}\label{diff_R_hat_tilde}
	& {\hat R} - {\tilde R} = \log \left( 1 + p_1 | h_1^{(n)} |^2 \right) + \log \left( 1 + p_2 | g_2 |^2 \right) \nonumber\\
	& - \log \left( 1 + p_1 | h_1^{(n)} |^2 + p_2 | h_2^{(n)} |^2 \right) \nonumber\\
	& =\! \log \!\left( 1 \!+\! p_2 | g_2 |^2 \right) \!-\! \log \!\left[ 1 \!+\! p_2 | h_2^{(n)} |^2 \!\!\left( 1 \!+\! p_1 | h_1^{(n)} |^2 \right)^{\!-1} \right]\!\!.\!\!\!\!
\end{align}
Note that $p_1$ varies in $[0, P]$ and $h_2^{(n)} |^2 ( 1 + p_1 | h_1^{(n)} |^2 )^{-1}$ decreases with $p_1$. Then, if $| g_2 |^2 \leq | h_2^{(n)} |^2 ( 1 + P | h_1^{(n)} |^2 )^{-1}$, we know from (\ref{diff_R_hat_tilde}) that
\begin{equation}\label{case1}
	{\hat R} \leq {\tilde R}, ~\forall~ p_1 \in [0, P].
\end{equation}
In this case, problem (\ref{max_SR_EJ2}) reduces to (\ref{max_R_hat}) with $({\hat P}_{1, {\text {lb}}}, {\hat P}_{1, {\text {ub}}}) = (0, P)$, and can be optimally solved based on (\ref{optimal_p1}).

If $| g_2 |^2 \geq | h_2^{(n)} |^2$, we know from (\ref{diff_R_hat_tilde}) that
\begin{equation}\label{case2}
	{\hat R} \geq {\tilde R}, ~\forall~ p_1 \in [0, P].
\end{equation}
In this case, problem (\ref{max_SR_EJ2}) reduces to (\ref{max_R_tilde}) with $({\tilde P}_{1, {\text {lb}}}, {\tilde P}_{1, {\text {ub}}}) = (0, P)$, and the optimal solution can be obtained from (\ref{opt_power_R_tilde}).

If $| h_2^{(n)} |^2 ( 1 + P | h_1^{(n)} |^2 )^{-1} < | g_2 |^2 < | h_2^{(n)} |^2$, we can always find $\beta = (| h_2^{(n)} |^2 / | g_2 |^2 - 1) / | h_1^{(n)} |^2$ in $[0, P]$ such that $| g_2 |^2 = | h_2^{(n)} |^2 ( 1 + \beta | h_1^{(n)} |^2 )^{-1}$. Then, according to (\ref{diff_R_hat_tilde}), we have
\begin{equation}\label{case3}
	{\hat R} \left\{\!\!\!\!\!\!\!
	\begin{array}{ll}
		& \leq {\tilde R}, ~\forall p_1 \in [0, \beta], \vspace{0.5 em}\\
		& \geq {\tilde R}, ~\forall p_1 \in [\beta, P].
	\end{array} \right.
\end{equation}
Problem (\ref{max_SR_EJ2}) can thus be divided into two subproblems, i.e., (\ref{max_R_hat}) with $({\hat P}_{1, {\text {lb}}}, {\hat P}_{1, {\text {ub}}}) = (0, \beta)$ and (\ref{max_R_tilde}) with $({\tilde P}_{1, {\text {lb}}}, {\tilde P}_{1, {\text {ub}}}) = (\beta, P)$. Both of them can be optimally solved based on Theorem~\ref{theorem1}. Theorem~\ref{theorem2} is thus proven.

\bibliographystyle{IEEEtran}

\end{document}